\documentclass[12pt,amsmath]{iopart}

\usepackage{graphicx}

\begin{document}

\title{Tracking anisotropic scattering in overdoped Tl$_2$Ba$_2$CuO$_{6+\delta}$ above 100 K}

\author{M M J French$^1$, J G Analytis$^1$, A Carrington$^1$, L Balicas$^2$ and N E Hussey$^1$}
\address{$^1$ H. H. Wills Physics Laboratory, University of Bristol, Tyndall Avenue, Bristol, BS8 1TL, UK}
\address{$^2$National High Magnetic Field Laboratory, Florida State University, Tallahassee, FL-32306, USA}

\begin{abstract}
This article describes new polar angle-dependent magnetoresistance (ADMR) measurements in the overdoped cuprate
Tl$_2$Ba$_2$CuO$_{6+\delta}$ over an expanded range of temperatures and azimuthal angles. These detailed measurements
re-affirm the analysis of earlier data taken over a more restricted temperature range and at a single azimuthal
orientation, in particular the delineation of the intraplane scattering rate into isotropic and anisotropic components.
These new measurements also reveal additional features in the temperature and momentum dependence of the scattering
rate, including anisotropy in the $T^2$ component and the preservation of both the $T$-linear and $T^2$ components up
to 100 K. The resultant form of the scattering rate places firm constraints on the development of any forthcoming
theoretical framework for the normal state charge response of high temperature superconducting cuprates.
\end{abstract}




\submitto{\NJP}

\maketitle

\section{Introduction}

Accepting the adage \lq what scatters may also pair', one might consider electrical resistivity a simple, albeit crude,
experimental probe of the pairing mechanism in a superconductor. Certainly in conventional BCS superconductors, the
strength of the electron-phonon (e-ph) coupling is reflected in the magnitude of the normal state resistivity, or more
precisely, the strength of the transport scattering rate 1/$\tau$. At elevated temperatures, where the resistivity is
$T$-linear, 1/$\tau$ is directly proportional to the transport e-ph coupling constant $\lambda_{\rm e-ph}$. The
superconducting transition temperature $T_c$  on the other hand is parameterized by the e-ph mass enhancement factor
$\lambda$ via the BCS or McMillan formula. In elemental metals, $\lambda$ and $\lambda_{\rm e-ph}$ are comparable
\cite{Sanborn89} though this is not always the case.

In high-$T_c$ superconducting cuprates (HTSC), a robust $T$-linear in-plane resistivity $\rho_{ab}(T)$, extending in
some cases up to 1000 K \cite{GurvitchFiory87}, is manifest only at or around optimal doping \cite{Ando04}.  In this
sense, just as in conventional superconductors, there appears an intimate correlation between the normal state
resistivity and the superconductivity. The high values of $\rho_{ab}(T)$, the $d$-wave pairing symmetry, proximity to
the antiferromagnetic Mott insulating state and the distinct power laws of $\rho_{ab}(T)$ and the inverse Hall angle
${\rm cot}\theta_{\rm H}(T)$ \cite{Hussey08} however all point towards a non-phononic origin of the $T$-linear
resistivity in HTSC. And whilst many proposals have been put forward, the origin of this and the other anomalous
transport properties in HTSC remains a profound theoretical challenge.

Alongside certain spectroscopic probes that have been applied to address the riddle of the normal state transport in
HTSC, including angle-resolved photoemission spectroscopy (ARPES), Raman scattering and optical conductivity,
angle-dependent magnetoresistance (ADMR) has emerged in recent years as a powerful ally. In a strong magnetic field $H$
(or more precisely, when the product of the cyclotron frequency and scattering time $\omega_c\tau \sim 1$), the
interlayer resistivity $\rho_{\perp}$ of a quasi-two-dimensional (Q2D) metal oscillates as $H$ is rotated in a polar
plane relative to the conducting layers \cite{Yamaji}. Maxima in $\rho_{\perp}$ occur when the velocity perpendicular
to the layers $v_{\perp}$, averaged over its trajectories on the Fermi surface (FS), is zero. Measurements of polar
ADMR at different azimuthal angles $\varphi$ were used originally to determine the size and shape of FS of various
anisotropic (layered) metals with long mean-free-paths at very low temperatures~\cite{Kartsovnik04, Bergemann03}.
Recent advances in both magnetic field technology and analytical techniques however have widened the scope of ADMR to
encompass less pure, correlated metals with shorter mean-free-paths \cite{Hussey03a, Balicas05}, weakly incoherent
metals \cite{McKenzie98} and with specific reference to the cuprates, anisotropic scattering rates \cite{Majed06,
Kennett07, Analytis07}.

ADMR should not be confused with Shubnikov-de Haas (SdH) magnetoresistance oscillations, which occur when the
magnitude, and not the orientation, of the magnetic field is changed. Unlike Shubnikov-de Haas oscillations, the
observation of ADMR is not impaired by the thermal damping of the quasiparticles (other than through the reduction in
$\omega_c\tau$). Thus, provided that $\omega_c \tau$ is large enough, ADMR measurements can be made over a much broader
temperature range than SdH.

The introduction of a $T$-dependent scattering rate anisotropy into the analysis of ADMR, had previously been applied
successfully to an overdoped Tl$_2$Ba$_2$CuO$_{6+\delta}$ (Tl2201) single crystal ($T_c \sim 15$ K) at one fixed
azimuthal angle over a temperature range 4.2 K $\leq T \leq$ 55 K \cite{Majed06}. These measurements revealed for the
first time different components of the in-plane transport scattering rate $\Gamma$($T$,{\bf k}), one isotropic and
quadratic in $T$, the other anisotropic, maximal near the saddle points at ($\pi$, 0) and proportional to $T$
\cite{Majed06}. Significantly, the deduced form of the scattering rate provided a consistent description of the
in-plane resistivity and Hall effect, albeit over a relatively narrow temperature range.

In this contribution, we describe further ADMR measurements in heavily overdoped Tl2201 over an expanded range of
temperatures (up to 110 K) and at several azimuthal angles. These detailed measurements demonstrate clearly that the
conclusions of those earlier measurements were robust and that the delineation of the intraplane scattering rate in
Tl2201 into these two distinct components, with only minor modifications, is entirely appropriate. We discuss possible
origins of the anomalous, anisotropic $T$-linear scattering rate and its possible relevance to high temperature
superconductivity.

\section{Experiment and Results}

Three self-flux grown crystals \cite{Tylerthesis} (typical dimensions 0.3 $\times$ 0.3 $\times$ 0.03mm$^3$) were
annealed at temperatures 300 - 360$^o$C in flowing O$_2$ and mounted in a $c$-axis quasi-Montgomery configuration. ADMR
were measured on a two-axis rotator in the 45 Tesla hybrid magnet at the NHMFL in Florida using a conventional
four-probe ac lock-in technique. The orientation of the crystal faces was indexed for selected crystals using a single
crystal X-ray diffractometer.

\begin{figure}
\begin{center}
\includegraphics [width=12.5cm, bb=0 0 3316 4944]{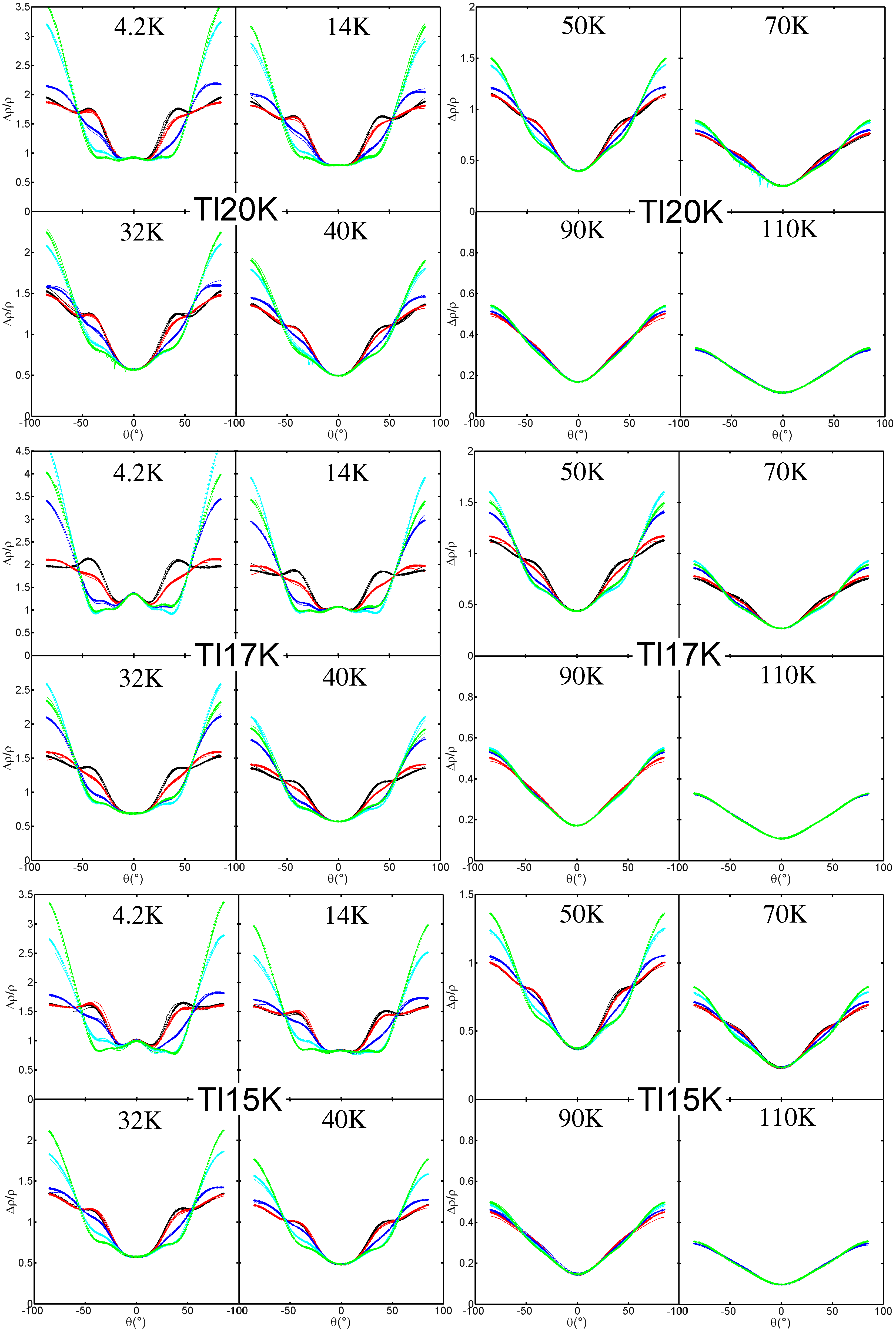}
\caption {Temperature dependent ADMR measurements at $\mu_0H$ = 45 Tesla on three overdoped Tl2201 single crystals with comparable $T_c$
values. The coloured circles are the actual polar ADMR data points, the different colours representing the different azimuthal angles relative to the Cu-O-Cu bond direction at which the data were taken. Tl15K - black $\phi=-3^\circ$, red $\phi=7^\circ$, dark blue $\phi=20^\circ$, light blue $\phi=33^\circ$ and green$\phi=43^\circ$. Tl20K - black $\phi=-2^\circ$, red $\phi=8^\circ$, dark blue $\phi=21^\circ$, light blue $\phi=34^\circ$ and green $\phi=44^\circ$. Tl17K - black $\phi=8^\circ$, red $\phi=18^\circ$, dark blue $\phi=32^\circ$, light blue $\phi=45^\circ$ and green $\phi=55^\circ$. The thin solid lines represent the fits to the Shockley-Chambers tube integral formalism incorporating anisotropic scattering.}\label{Figure1}
\end{center}
\end{figure}

Figure \ref{Figure1} shows polar ADMR data, plotted as $\Delta\rho_\perp/\rho_\perp(0)$ (i.e. normalised to their
zero-field value), for three different crystals, Tl15K, Tl17K and Tl20K (the numbers referring to their respective
$T_c$ values) taken at up to 5 different azimuthal angles $\varphi$ (relative to the Cu-O-Cu bond direction) and
different temperatures up to 110 K. All data were obtained in an applied magnetic field of 45 Tesla. The coloured
circles are the actual data whilst the thin (black) solid lines represent fits to the data using the formalism
described below. The small asymmetry around $\theta$=0 arises from slight misalignment of the crystalline $c$-axis with
respect to the rotational platform which is incorporated into our model with negligible impact on the parameters
obtained.

The reproducibility of the ADMR response in Tl2201 for different crystals is demonstrably illustrated in this series of
panels. Clear azimuthal dependence of the polar ADMR sweeps can be seen in all crystals up to 90 K, and is resolvable,
though not visible perhaps in this Figure, at $T$ = 110 K. To our knowledge, this is the highest recorded temperature
at which ADMR measurements have been reported and highlights the unique advantage of this technique over SdH
oscillations in allowing one to extract from the data useful information on the evolution of physical parameters with
temperature. The 100 K landmark is also significant that it demonstrates that ADMR measurements may eventually be
observed in optimally doped Tl2201 ($T_c \sim$ 90 K). Finally, as will be discussed in more detail below, the quality
of the fits for what is a fixed (i.e. $T$-independent) FS parameterization illustrates how robustly one can track the
evolution of other relevant $T$-dependent quantities from ADMR.

\section{Fermi-surface parameterization}

In order to analyse the data, we carried out a least-square fitting using the Shockley-Chambers tube integral form of
the Boltzmann transport equation modified for a Q2D metal with a fourfold anisotropic scattering rate
$1/\tau(\psi)=(1+\alpha\cos 4\psi)/\tau_0$ and anisotropic in-plane velocity $v_F(\psi)$, incorporated via
$1/\omega_c(\psi)=(1+\beta\cos 4\psi)/\omega_0$ \cite{Majed06, Kennett07, Analytis07}. The sign of $\alpha$ ($\beta$)
defines the location of maximal scattering (density of states) respectively. The FS wave vector $k_F(\theta,\phi)$ is
parameterized by the lowest-order harmonic components satisfying the body-centered-tetragonal symmetry of
Tl2201~\cite{Bergemann03, Analytis07}

\begin{table}[t]
\begin{center}
%\begin{ruledtabular}
\begin{tabular}{ccccc}

Sample & $k_{00}~(\AA^{-1})$ & $k_{40}$ & $k_{61}/k_{21}$ & $k_{101}/k_{21}$ \\
\hline
Tl20K & 0.728 & 0.033 & 0.646 & -0.306\\
Tl17K & 0.729 & 0.034 & 0.698 & -0.258\\
Tl15K & 0.730 & 0.032 & 0.711 & -0.249\\
%NP15K & 0.733 & 0.049 & 0.68 & -0.20\\
\end{tabular}
%\end{ruledtabular}
\end{center}
\caption{\label{tab:table1} Fermi-surface parameters obtained from the least-squared fitting of 4.2 K ADMR data for the
three different Tl2201 crystals shown in Figure 1.} \vspace{0.3cm}
\end{table}

\begin{eqnarray}
&&k_F(\theta,\varphi) = k_{00}(1 - k_{40}\cos 4\varphi) - k_{21}\cos(k_zc/2)\sin 2\varphi \\
&&- k_{61}\cos(k_zc/2)\sin 6\varphi - k_{101}\cos(k_zc/2)\sin 10\varphi
\end{eqnarray}

\par
where $k_z$ is the $c$-axis wave vector and $c$ the interlayer spacing. Note that the $c$-axis warping parameters
$k_{21}$, $k_{61}$, and $k_{101}$ are small compared to $k_{00}$, the radius of the cylindrical FS (about the zone
corners), and $k_{40}$, its in-plane squareness, and only ratios (e.g., $k_{61}/k_{21}$) can be determined to good
accuracy. $\beta$ depends largely on our choice of $k_{61}/k_{21}$ with the best least-square values giving
$\beta=0\pm0.1$ for $0.6\leq k_{61}/k_{21} \leq 0.8$~\cite{Analytis07}. The sum $\alpha+\beta$ is much less sensitive
to variations in $k_{61}/k_{21}$ and for simplicity (and to reduce the number of free parameters), we assume hereafter
$\omega_c(\varphi)=\omega_0$. This has only a negligible effect on the quality of the fits or the other fitting
parameters. We also fix $k_{00}$ using the empirical relation $T_c/T_c^{\rm max}=1-82.6(p-0.16)^2$ with $T_c^{\rm
max}=92K$ and $(\pi k_{00}^2)/(2\pi/a)^2=(1+p)/2$~\cite{Presland} which agrees extremely well with values obtained from
recent de Haas-van Alphen (dHvA) and SdH measurements~\cite{Vignolle08}. The initial fit for each sample was performed
on the lowest temperature data (4.2K). The values for the parameters $k_{04}$, $k_{61}$ and $k_{101}$ from this initial
fit were then kept fixed for all elevated temperatures with only $\omega_0\tau_0$ and $\gamma$ as free parameters.

The best fits, shown overlayed on the data, are all excellent and the parameters (listed in Table 1) appear well
constrained due to the wide range of polar and azimuthal angles and temperatures studied. Moreover, the projected
in-plane FS is found to be in excellent agreement with a recent ARPES study on the same compound \cite{Plate05}.

\section{$T$-dependence of the anisotropic scattering}

\begin{figure}
\begin{center}
\includegraphics [width=14.0cm, bb=0 0 946 542]{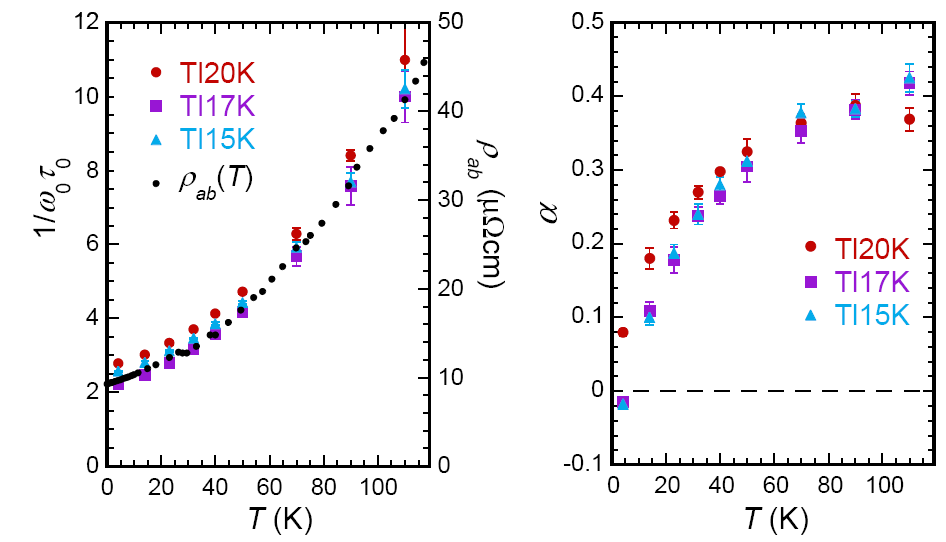}
\caption {Temperature dependence of 1/$\omega_0\tau_0$ and $\alpha$ obtained from the ADMR measurements shown in Figure
1. The black dots in the left-hand panel refer to resistivity data on a Tl2201 crystal with a similar $T_c$
\cite{Mackenzie96}. The $\rho_{ab}(T)$ values have been shifted by + 3 $\mu\Omega$cm to coincide with
1/$\omega_0\tau_0$. Note too that the superconductivity in this crystal has been suppressed by a magnetic field.}
\label{Figure2}
\end{center}
\end{figure}

It is clear from Figure 1 that the same FS parameterization can fit the experimental ADMR data in overdoped Tl2201 at
all azimuthal angles studied and at all temperatures up to 110 K. The addition of the $T$-dependent anisotropy factor
$\alpha$ is thus a simple and elegant way to track the evolution of the ADMR with increasing temperature. Figure 2
shows the $T$-dependence of $1/\omega_0\tau_0$ (left panel) and $\alpha$ (right panel) for the three crystals extracted
from the fits shown in Figure 1. Good consistency is found between all three crystals; $\alpha$ is found to increase
from a value close to zero, tending towards a constant value around 0.4 at high temperatures. The sign of $\alpha$
signifies that the maxima in both the linear and quadratic components of $1/\tau(\varphi)$ occur near the Brillouin
zone boundary where the superconducting gap (and the pseudogap in more underdoped cuprates) is maximal.

The value of 1/$\omega_0\tau_0$ at $T$ = 0 corresponds to $\omega_c\tau$ = 0.45 at $\mu_0H$ = 45 Tesla. Taking the
value for the effective mass ($m^* \sim 4.5 m_e$) of overdoped Tl2201 from recent quantum oscillation experiments
\cite{Vignolle08}, we obtain an estimate for the momentum-averaged zero-temperature impurity scattering rate
$\hbar/\tau(0) \simeq$ 3 meV. This is at least one order of magnitude less than that obtained for the single particle
scattering rate $\Gamma(0)$ from ARPES \cite{Plate05}. It is also significantly lower than the interplane hopping
integral $t_{\perp}$ estimated from the electrical resistivity anisotropy, implying that interplane transport in
overdoped Tl2201 is coherent, at least at low temperatures. Finally, as shown in the left-panel, the $T$-dependence of
$1/\omega_0\tau_0$ tracks that of the in-plane resistivity $\rho_{ab}(T)$, measured on a crystal from the same batch
\cite{Mackenzie96}, over the full temperature range. This observation confirms that the inclusion of $\alpha(T)$ in the
fitting procedure does not introduce any additional $T$-dependence in $1/\tau(T)$.

\begin{figure}
\begin{center}
\includegraphics [width=10.0cm, bb=0 0 592 999]{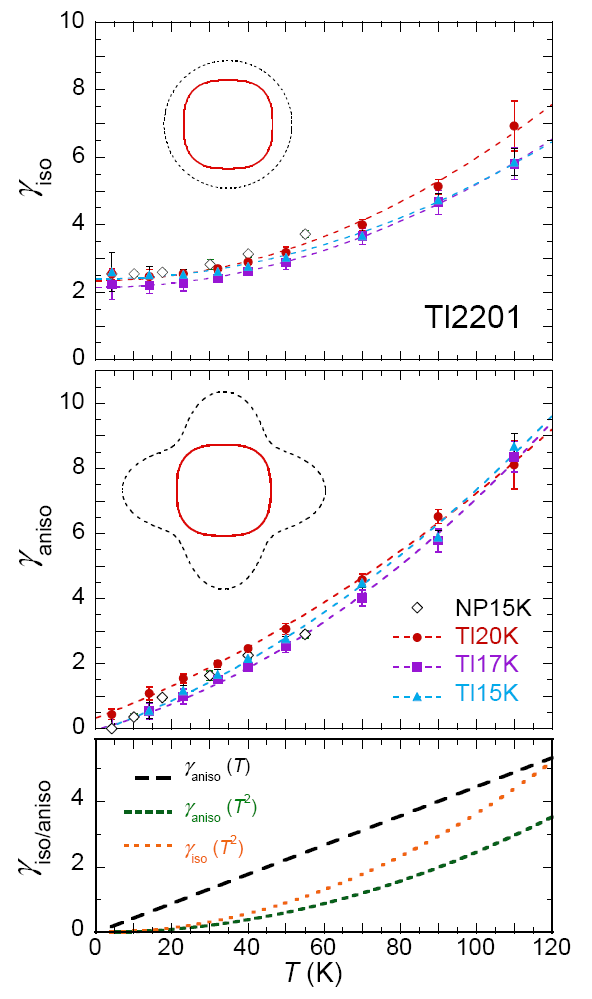}
\caption {Temperature dependence of the isotropic (top panel) and anisotropic (middle panel) scattering rates
determined from the ADMR measurements shown in Figure 1. NP15K refers to the sample whose ADMR were measured at a
single azimuthal angle \cite{Majed06}. The dashed lines in top and middle panels are fits to $A + CT^2$ and $A + BT +
CT^2$ respectively. The insets in each panel depict the FS (as red solid lines) and the corresponding scattering rates
(as black dashed lines. Bottom panel: Components of $\gamma_{\rm aniso}(T)$ (black long-dashed lines and red
short-dashed lines) and $\gamma_{\rm iso}(T)$ (orange dots) for Tl15K.}\label{Figure3}
\end{center}
\end{figure}

As done previously \cite{Majed06}, we use a simple trigonometric identity to split $1/\omega_c\tau(\varphi)$ into two
components, one without $\varphi$-dependence ($\gamma_{\rm iso}=(1-\alpha)/\omega_0\tau_0$) and one with ($\gamma_{\rm
aniso}=2\alpha/\omega_0\tau_0\times\cos^22\varphi$). The top two panels of Figure 3 show the resultant $T$-dependence
of $\gamma_{\rm iso}$ (top panel) and $\gamma_{\rm aniso}$ (middle panel) for the three crystals studied. Also shown
for comparison (open diamonds) are the corresponding values derived from the original ADMR measurements carried out at
a fixed azimuthal angle \cite{Majed06}. It is noteworthy that the parameters extracted from an individual polar sweep
are entirely consistent with those more tightly constrained by multiple sweeps performed at different $\varphi$. This
consistency demonstrates that once the size and topology of the FS have been determined at low $T$, subsequent
parameterization at any intermediate temperature can be obtained from a single polar angle sweep.

The distinct $T$-dependencies of $\gamma_{\rm iso}$ and $\gamma_{\rm aniso}$ are evident from this Figure. As indicated
by the dashed lines in the top panel, the isotropic component is well fitted by the expression $\gamma_{\rm iso} =
A+BT^2$ implying a combination of impurity and fermion-fermion scattering. The quadratic form of $\gamma_{\rm iso}(T)$
extends up to 100 K, a significant fraction of the Fermi energy in Tl2201 ($\sim$ 0.4 eV \cite{Vignolle08}). Whilst this may seem surprising, we note
that $T^2$ resistivities extending over tens of Kelvin have been reported in a number of correlated oxides, including
non-superconducting cuprates \cite{McBrien02, Nakamae03}, ruthenates \cite{Maeno97}, rhodates \cite{Perry06} and
titanates \cite{Tokura93}.

In contrast to $\gamma_{\rm iso}(T)$, $\gamma_{\rm aniso}(T)$ has a negligible intercept at $T$ = 0 K, implying that
there is no anisotropy in the impurity scattering rate, or rather, that the isotropic-$\ell$ approximation is obeyed in
overdoped Tl2201 at low temperatures. This contrasts with the case in overdoped LSCO where recent analysis of Hall
effect and ARPES revealed a striking ($> 3$) anisotropy in the impurity scattering rate \cite{Narduzzo08} attributed to
small-angle scattering off out-of-plane Sr substitutional disorder. \cite{Varma01}. In Tl2201, defects also reside
predominantly out of the plane, either as Cu substitution on the Tl site, or as interstitial oxygen. However, strong
next-nearest-neighbour hopping in Tl2201 gives rise to an approximately square-shaped FS (see Figure 3) which crosses
the Brillouin zone boundary far from the extended saddle point near ($\pi$, 0). In LSCO by contrast, $t'/t$ is smaller
and as a result, the FS is more diamond-shaped and for similar filling factors, the FS is pushed closer to the van Hove
singularity at ($\pi$, 0). As a result, the basal plane anisotropy in the density of states in LSCO \cite{Yoshida07} is
almost one order of magnitude larger than in Tl2201 \cite{Plate05}, leading to a very different anisotropy in
$\gamma_{\rm aniso}(0)$.

Below 50 K, $\gamma_{\rm aniso}$ is dominated by a robust $T$-linear component. At higher $T$, $\gamma_{\rm aniso}(T)$
develops slight upward curvature. A fit to a second order polynomial provides a good description of $\gamma_{\rm
aniso}(T)$ over the full temperature range studied, as indicated by the dashed lines in the lower panel. This
additional $T^2$ term in $\gamma_{\rm aniso}(T)$ was not picked out in the original lower temperature measurements
\cite{Majed06} as its contribution to $\gamma_{\rm aniso}(T)$ was too small to be significant. The relative magnitudes
of the two components of $\gamma_{\rm aniso}(T)$ are plotted for Tl15K in the bottom panel of Figure 3 along with
$\gamma_{\rm iso}(T)$. There is roughly 60\% anisotropy in the $T^2$ scattering rate within the basal plane, comparable
to the variation in the density of states \cite{Plate05}. Whilst finite temperature effects (e.g. effects due to
broadening of the Fermi distribution function) not included in our Boltzmann analysis may affect the form of
$\gamma_{\rm aniso}(T)$ and $\gamma_{\rm iso}(T)$ at the highest temperatures \cite{Brinkmann08}, it is worth pointing
out that this parameterization provides a consistent description of the transport properties up to 110 K, as discussed
in the following section.

It is interesting to compare this form of the transport scattering rate with the single particle scattering rate
$\Gamma(\omega)$ measured by ARPES. In overdoped and near-optimally doped Bi$_2$Sr$_2$CaCu$_2$O$_8$ (Bi2212), for
example, $\Gamma(\omega)$ along the nodal directions is approximately quadratic over a broad energy range
\cite{Kordyuk04, Koralek06}. Coupled with the quadratic temperature dependence of $\gamma_{\rm iso}(T)$, the sole
contribution along the zone diagonals, this appears to identify electron-electron (Umklapp) scattering as the principal
source of damping for nodal quasiparticles. Away from ($\pi, \pi$), $\Gamma(\omega)$ develops an additional
$\omega$-linear component \cite{Kordyuk04}. In LSCO, this $\omega$-linear term becomes dominant away from the nodes,
with a magnitude that displays a cos2$\varphi$ dependence reminiscent of what is seen in $\gamma_{\rm aniso}(\varphi)$
\cite{Chang08}. Thus, both $T$-dependent components of 1/$\tau$ appear to have a frequency-dependent counterpart in the
single particle channel.

\section{Comparison with in-plane transport properties}

\begin{figure}
\begin{center}
\includegraphics [width=9.0cm, bb=0 0 519 783]{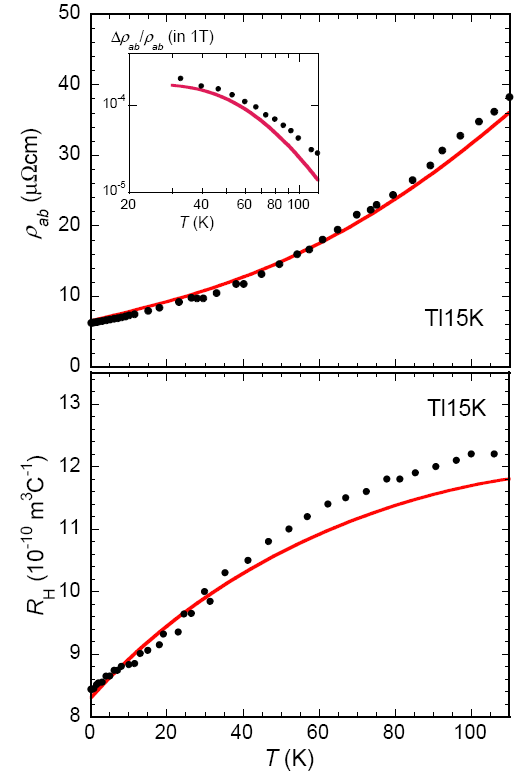}
\caption {Temperature dependence of the in-plane resistivity (top panel) and in-plane Hall coefficient (bottom panel)
for overdoped Tl2201 ($T_c$ = 15 K). The black circles are the experimental data are taken from Ref. \cite{Mackenzie96}
whilst the solid red lines are simulations for Tl15K obtained from the FS parameters displayed in Table 1 and the
parameters plotted in Figure 2 (displaced by - 3 $\mu\Omega$cm). Inset: Corresponding plot of the in-plane
magnetoresistance $\Delta\rho_{ab}/\rho_{ab}(T)$ on a log-log scale. Experimental data shown here are for a $T_c$ = 25
K sample reported in Ref. \cite{Hussey96}.} \label{Figure4}
\end{center}
\end{figure}

Armed with full FS information and a quantitative determination of the momentum and temperature dependence of the
in-plane mean-free-path (via $1/\omega_c\tau$($\varphi$, $T$)), one can in principle calculate any in-plane transport
property since the individual coefficients of the in-plane conductivity tensor \cite{Hussey03b} can be re-formulated
solely in terms of the parameters derived from ADMR. For example, the in-plane resistivity and Hall conductivity can be
expressed as:

\begin{eqnarray}
1/\rho_{ab}(T) = \sigma_{xx}(T) = \frac{e}{2\pi^2 c} \int \frac{k_F^2(\varphi)}{1/\omega_c\tau(\varphi, T)} d\varphi \\
\\
\sigma_{xy}(T) = \frac{e}{2\pi^2 c} \int \frac{k_F^2(\varphi)}{1/\omega_c\tau(\varphi, T)} \frac{\partial
(\omega_c\tau(\varphi, T))}{\partial \varphi} d\varphi
\end{eqnarray}

\par
where $1/\omega_c\tau(\varphi, T)$ is given by its value at $\mu_0H$ = 1 Tesla. It is important to realize at this
stage that inserting $1/\omega_c\tau(\varphi,T)=(1+\alpha\cos 4\varphi)/\omega_0\tau_0(T)$ into the above expressions
would give {\it precisely} the same result as if one used $\gamma_{\rm iso}(T)$ + $\gamma_{\rm aniso}(T)$. Hence, for
the purposes of simulating the data, the two combinations are equivalent - the latter is being considered the more
physical only because both terms can be deconvolved into individual components each obeying an integer power-law
$T$-dependence.

Figure 4a shows $\rho_{ab}(T)$ as determined for Tl15K from our ADMR analysis, superimposed on the same resistivity
data depicted in Figure 2 \cite{Mackenzie96}. The form of $\rho_{ab}(T)$, in particular the strong $T$-linear component
below 40K and the development of supra-linear behaviour above this temperature, is well reproduced by the
parameterization. Comparison with Figure 3 reveals that $\rho_{ab}(T)$ is largely determined by $\gamma_{\rm
aniso}(T)$.

The corresponding in-plane Hall coefficient $R_{\rm H}(T)$ is shown in Figure 4b. The calculated value of $R_{\rm H}$
at $T$ = 0 is found to be in excellent agreement with the experimental data, confirming that in heavily overdoped
Tl2201, the isotropic-$\ell$ approximation is a good one and that $R_{\rm H}$(0) can be estimated directly from the
size of the FS. The absolute change in anisotropy in 1/$\omega_c\tau(\varphi, T)$ can also account fully for the
significant (50\%) rise in $R_{\rm H}(T)$ between 0 K and 110 K. Finally, the magnetoresistance
$\Delta\rho_{ab}/\rho_{ab}$ (inset in Figure 4a) is seen to have the correct magnitude and $T$-dependence
\cite{Hussey96}. Deviations between the data and the simulation are larger here, as one might expect for a second-order
galvanometric effect. Overall, the parameterization described in Figure 3 gives an excellent account, not only of the
evolution of the ADMR signal (Figure 1), but also of the various transport coefficients at this elevated doping level
($p \sim 0.26)$. Analysis of other doping concentrations with higher $T_c$ values suggests that discrepancies begin to
appear as $T_c$ increases \cite{Majed07}, possibly due to the emergence of vertex corrections \cite{Kontani06} that
manifest themselves only in the in-plane transport \cite{Sandemann01} and are thus transparent to ADMR.

\section{Discussion}

The above analysis provides strong evidence that the anomalous transport properties in HTSC are, to a large degree,
governed by $T$-dependent anisotropy in the transport scattering rate. Such anisotropy has been attributed over time to
a host of possible scattering mechanisms, such as spin fluctuations \cite{Carrington92, MonthouxPines}, charge
fluctuations \cite{Castellani95}, anisotropic e-e (Umklapp) scattering \cite{Hussey03b, Hussey06} or $d$-wave
superconducting fluctuations \cite{IoffeMillis}.

In many of these scenarios, one considers a single dominant scattering process having distinct $T$-dependencies at
different regions in $k$-space. In such a case, it is necessary to sum the different $\tau$, rather than $1/\tau$, in
any integration around the in-plane FS. The fact then that $1/\tau$ in overdoped Tl2201 can be delineated into two {\it
additive} components implies that two independent quasiparticle scattering processes must coexist {\it everywhere} on
the cuprate FS, except perhaps along the zone diagonals. This claim is supported by recent resistivity measurements on
LSCO in which $\rho_{ab}(T)$ is seen to be composed of two series-connected $T$-dependent components each with an
integer exponents (i.e. $\rho_{ab}(T)$ = $\alpha_0 + \alpha_1 T + \alpha T^2$) \cite{Cooper08}.

The same resistivity study revealed two other notable features in the doping/temperature phase diagram; namely the
persistence of the $\alpha_1 T$ term, both down to very low temperatures (of order 1 K) and over a broad range of
doping, from optimal doping to the edge of the superconducting dome on the overdoped side \cite{Cooper08}.
Collectively, these numerous features (the additive components to 1/$\tau$, the form of the anisotropy in $\gamma_{\rm
aniso}$, the vanishing of the $T$-linear component along the nodes, its persistence to low temperatures and across the
overdoped regime and its correlation with $T_c$) place very strong constraints on any theoretical model put forward to
explain the charge dynamics of HTSC. Indeed, to the best of our knowledge, no model had been able to predict the
precise form of 1/$\omega_c\tau(\varphi,T)$ {\it prior} to publication of Ref. \cite{Majed06}. And whilst it is true
that detailed information on 1/$\omega_c\tau(\varphi,T)$ has only been obtained thus far on Tl2201 single crystals over
a very narrow doping range close to the superconductor/metal boundary, the gradual evolution of the transport
properties in both Tl2201 and LSCO with doping \cite{Majed07, Cooper08, Manako92} suggests strongly that similar
phenomenology may exist across the entire overdoped region of the cuprate phase diagram.

We conclude by highlighting four independent theoretical approaches that have managed to capture some, though not all,
of the salient experimental features. As discussed above, the linear component of $\gamma_{\rm aniso}(T)$ and its
angular dependence are mirrored in $\Gamma(\omega)$ \cite{Chang08}, suggesting that the two derive from the same
origin. This combined linearity in both the temperature and frequency scales is typically referred to as marginal
Fermi-liquid (MFL) phenomenology \cite{Varma89}. In its original guise \cite{Varma89}, the MFL self-energy was
considered to have no momentum dependence. A recent microscopic model \cite{Aji07} however has shown that the
fluctuations that lead to MFL physics can generate a factor of two anisotropy in $\Gamma$ \cite{Zhu08}. A robust
$T$-linear scattering rate is also expected in a 2D boson-fermion mixture  \cite{Alexandrov97, Wilson08} due to fermion
scattering off density fluctuations of charged 2$e$ bosons \cite{Alexandrov97}. The $T$-linearity in this case arises
from the fact that the screening radius squared for direct Coulomb repulsion between free fermions and heavy, almost
localized bosons is proportional to temperature above the condensation temperature $T_B$. This model thus predicts that
the $T$-linear scattering rate should be manifest in the normal state at all temperatures above $T_B(H)$.

By assuming an effective interaction with the appropriate $d$-wave form factor, the transport scattering rate and
electrical resistivity of a 2D metal close to a Pomeranchuk instability was recently shown to follow a similar
linear-in-$T$ dependence as $T \rightarrow 0$ \cite{Dell'Anna07}. When combined with impurity scattering and
conventional, isotropic electron-electron scattering, this gives rise to a scattering rate (and resistivity) with a
form identical to that observed in overdoped Tl2201. It is not yet clear however why the strength of the $T$-linear
scattering rate within this scenario should scale with $T_c$.  Finally, in recent functional renormalization group
calculations for a 2D Hubbard model, Ossadnik and co-workers have found a strongly angle-dependent $T$-linear
scattering term which derives from a scattering vertex that increases as the energy (temperature) scale is lowered
\cite{Ossadnik08}. Significantly, this $T$-linear scattering rate shows strong doping dependence too and vanishes as
the superconductivity disappears on the overdoped side, consistent with experimental observations \cite{Majed07,
Cooper08}.

Despite these encouraging developments, it will be some time before all features of the scattering rate phenomenology
uncovered by ADMR and the other high-field transport experiments are contained within a single theoretical framework,
but its apparent correlation with high temperature superconductivity, both in terms of its magnitude and its
anisotropy, provides strong motivation to develop such a framework. As quoted at the beginning of this contribution,
\lq what scatters may also pair'. What is now clear in the case of HTSC, is that identification of the anomalous
scatterer will prove a key step in elucidating the mechanism of high temperature superconductivity itself.

\section{Conclusions}

In conclusion, we have performed a series of detailed ADMR measurements on three overdoped Tl2201 single crystals over
an extended temperature range up to 110 K. Consistent fitting of the data is achieved over the full temperature range
with a single FS parameterization and an anisotropic 1/$\tau$ whose anisotropy is found to vary in a manner that can
account for the $T$-dependence of the in-plane Hall effect over the same temperature range. This detailed study
re-affirms the analysis of earlier ADMR data taken at a single azimuthal angle over a more limited temperature range
and highlights the power of the ADMR technique in giving robust physical parameters that are currently inaccessible to
any other known experimental probe. In so doing, this study aptly demonstrates the unique ability of ADMR to access
hitherto hidden details of the charge response of correlated layered metals.

\ack We would like to thank A. P. Mackenzie for the single crystal samples and J. P. A. Charmant for assistance with
the X-ray diffraction analysis. This work was supported by the EPSRC (UK) and a co-operative agreement between the
State of Florida and NSF.


\section*{References}

\begin{thebibliography}{32}

\bibitem{Sanborn89} Sanborn B A, Allen P B and Papaconstantopoulos D A 1989 {\it Phys. Rev. B} {\bf 40} 6037
\bibitem{GurvitchFiory87} Gurvitch M and Fiory A T 1987 {\it Phys. Rev. Lett.} {\bf 59} 1337
\bibitem{Ando04} Ando Y, Komiya S, Segawa K, Ono S and Kurita Y 2004 {\it Phys. Rev. Lett.} {\bf 93} 267001
\bibitem{Hussey08} Hussey N E 2008 {\it J. Phys. - Condens. Matt} {\bf 20} 123201
\bibitem{Yamaji} Yamaji K 1989 {\it J. Phys. Soc. Japan} {\bf 58} 1520
\bibitem{Kartsovnik04} Kartsovnik M 2004 {\it Chem. Rev.} {\bf 104} 5737
\bibitem{Bergemann03} Bergemann C, Mackenzie A P, Julian S R, Forsythe D and Ohmichi E 2003 {\it Adv. Phys.} {\bf 52} 639
\bibitem{Hussey03a} Hussey N E, Abdel-Jawad M, Carrington A, Mackenzie A P and Balicas L 2003 {\it Nature} {\bf 425} 814
\bibitem{Balicas05} Balicas L, Abdel-Jawad M, Hussey N E, Chou F C and Lee P A 2005 {\it Phys. Rev. Lett.} {\bf 94} 236402
\bibitem{McKenzie98} McKenzie R H and Moses P 1998 {\it Phys. Rev. Lett.} {\bf 81} 4492
\bibitem{Majed06} Abdel-Jawad M, Kennett M P, Balicas L, Carrington A, Mackenzie A P, McKenzie R H and Hussey N E  2006 {\it Nature Phys.} {\bf 2}
821
\bibitem{Kennett07} Kennett M P and McKenzie R H 2007 {\it Phys. Rev. B} {\bf 76} 054515
\bibitem{Analytis07} Analytis J G, Abdel-Jawad M, Balicas L, French M M J and Hussey N E  2007 {\it Phys. Rev. B} {\bf 76} 104523
\bibitem{Tylerthesis} Tyler A W 1998 DPhil. thesis (Cambridge University)
\bibitem{Presland} Presland M R, \emph{et al.}, (1991) {\it Physica C} (Amsterdam) \textbf{176} 95
\bibitem{Vignolle08} Vignolle B, Carrington A, Cooper R A, French M M J, Mackenzie A P, Jaudet C, Vignolles D, Proust C and Hussey N E 2008
{\it Nature} {\bf 455} 952
\bibitem{Plate05} M. Plat$\acute{e}$ {\it et al.}, 2005 {\it Phys. Rev. Lett.} {\bf 95} 077001
\bibitem{Mackenzie96} Mackenzie A P, Julian S R, Sinclair D C and Lin C T 1996 {\it Phys. Rev. B} {\bf 53} 5848
\bibitem{McBrien02} McBrien M N, Hussey N E, Meeson P J, Horii S and Ikuta H 2002 {\it J. Phys. Soc. Japan} {\bf 71} 701
\bibitem{Nakamae03} Nakamae S, Behnia K, Yates S J C, Mangkorntong N, Nohara M, Takagi H and Hussey N E 2003 {\it Phys. Rev. B} {\bf 68} 100502
\bibitem{Maeno97} Maeno Y {\it et al.} 1997 {\it J. Phys. Soc. Japan} {\bf 66} 1405
\bibitem{Perry06} Perry R S, Baumberger F, Balicas L, Kikugawa N, Ingle N J C, Rost A, Mercure J F, Maeno Y, Shen Z X and
Mackenzie A P 2006 {\it New J. Phys.} {\bf 8} 175
\bibitem{Tokura93} Tokura Y, Taguchi Y, Okada Y, Fujishima Y, Arima T, Kumagai K and Iye Y 1993 {\it Phys. Rev. Lett.} {\bf 70} 2126
\bibitem{Narduzzo08} Narduzzo A, Albert G, French M M J, Mangkorntong N, Nohara M, Takagi H and Hussey N E 2008 {\it Phys. Rev. B} {\bf 77} 220502
\bibitem{Varma01} Varma C M and Abrahams E 2001 {\it Phys. Rev. Lett.} {\bf 86} 4652
\bibitem{Yoshida07} Yoshida T, Zhou X J, Lu D H, Komiya S, Ando Y, Eisaki H, Kakeshita T, Uchida S, Hussain Z, Shen Z-X and Fujimori A 2007
{\it J. Phys.: Condens. Matter} {\bf 19} 125209
\bibitem{Brinkmann08} Brinkmann B A M and Kennett M P 2008 cond-mat/0811.4442v1. In this work, the authors confirm that including finite
temperature effects do not affect the conclusion that there is an anisotropic contribution to the scattering rate in Tl2201
\bibitem{Kordyuk04} Kordyuk A A, Borisenko S V, Koitzsch A, Fink J, Knupfer M, Buchner B, Berger H, Margaritondo G, Lin C T, Keimer B, Ono S and Ando Y 2004 {\it Phys. Rev. Lett.} {\bf 92} 257006
\bibitem{Koralek06} Koralek J D, Douglas J F, Plumb N C, Sun Z, Fedorov A V, Murnane M M, Kapteyn H C, Cundiff S T, Aiura Y, Oka K, Eisaki H and Dessau D S 2006 {\it Phys. Rev. Lett.} {\bf 96} 017005
\bibitem{Chang08} Chang J, Shi M, Pailhés S, Månsson M, ClaessonT, Tjernberg O, Bendounan A, Sassa Y, Patthey L, Momono N, Oda M, Ido M, Guerrero S, Mudry C and Mesot J 2008 {\it Phys. Rev. B} {\bf 78} 205103
\bibitem{Hussey03b} Hussey N E 2003 {\it Euro. Phys. J. B} {\bf 31} 495
\bibitem{Hussey96} Hussey N E, Cooper J R, Wheatley J M, Fisher I R, Carrington A, Mackenzie A P, Lin C T and Milat O 1996 {\it Phys. Rev. Lett.} {\bf 76} 122
\bibitem{Majed07} Abdel-Jawad M, Analytis J G, Balicas L, Carrington A, Charmany J P A, French M M J and Hussey N E  2007 {\it Phys. Rev. Lett.} {\bf 99} 107002
\bibitem{Kontani06} Kontani H 2006 {\it J. Phys. Soc. Japan} {\bf 75} 013703
\bibitem{Sandemann01} Sandemann K G and Schofield A J 2001 {\it Phys. Rev. B} {63} 094510
\bibitem{Carrington92} Carrington A, Mackenzie A P, Lin C T and Cooper J R 1992 {\it Phys. Rev. Lett.} {\bf 69} 2855
\bibitem{MonthouxPines} Monthoux P and Pines D 1992 {\it Phys. Rev. B}{\bf 49} 4261
\bibitem{Castellani95} Castellani C, di Castro C and Grilli M 1995 {\it Phys. Rev. Lett.} {\bf 75} 4650
\bibitem{Hussey06} Hussey N E, Alexander J C and Cooper R A 2006 {\it Phys. Rev. B} {\bf 74} 214508.1-8.
\bibitem{IoffeMillis} Ioffe L B and Millis A J 1998 {\it Phys. Rev. B} {\bf 58} 11631
\bibitem{Cooper08} Cooper R A, Wang Y, Vignolle B, Lipscombe O J, Hayden S M, Tanabe Y, Adachi T, Koike Y, Nohara M, Takagi H, Proust C and
Hussey N E 2009 {\it Science} {\bf 323} 603
\bibitem{Manako92} Manako T, Kubo Y and Shimakawa T 1992 {\it Phys. Rev. B} {\bf 46} 11019
\bibitem{Varma89} Varma C M, Littlewood P B, Schmitt-Rink S, Abrahams E and Ruckenstein A E 1989 {\it Phys. Rev. Lett.} {\bf 63} 1996
\bibitem{Aji07} Aji V and Varma C M 2007 {\it Phys. Rev. Lett.} {\bf 63} 1996
\bibitem{Zhu08} Zhu L, Aji V, Shekhter A and Varma C M 2008 {\it Phys. Rev. Lett.} {\bf 100} 057001
\bibitem{Alexandrov97} Alexandrov A S 1997 {\it Physica} {\bf 274C} 237
\bibitem{Wilson08} Wilson J A 2008 arXiV:condmat0811.3096v1
\bibitem{Dell'Anna07} Dell'Anna L and Metzner W 2007 {\it Phys. Rev. Lett.} {\bf 98} 136402
\bibitem{Ossadnik08} Ossadnik M, Honerkamp C, Rice T M and Sigrist M 2008 {\it Phys. Rev. Lett.} {\bf 101} 256405

\end{thebibliography}
\end{document}